# Room-temperature operation of Si spin MOSFET with high on/off spin signal ratio

Takayuki Tahara [1], Hayato Koike [2], Makoto Kameno [1], Yuichiro Ando [1], Kazuhito Tanaka [3], Shinji Miwa [3], Yoshishige Suzuki [3] and Masashi Shiraishi [1,$]

1. Department of Electronic Science and Engineering, Kyoto University, Kyoto 615-8510, Japan.
2. Tchnology HQ, TDK Corporation, Chiba 272-8558, Japan.
3. Graduate School of Engineering Science, Osaka University, Toyonaka 560-8531, Japan.

$ Corresponding author (mshiraishi@kuee.kyoto-u.ac.jp)

**Experimental demonstration of a Si spin metal-oxide-semiconductor field-effect transistor (MOSFET) with high on/off ratio of the source-drain current and spin signals at room temperature is reported. Non-degenerate n-type Si is used as a spin channel, and an effective application of the gate voltage in a back gate structure enables the demonstration of the spin MOSFET. This achievement can pave a way to a practical application of a Si spin MOSFET.**

Semiconductor spintronics has been intensively studied in several decades. In the earlier stage of semiconductor spintronics, GaAs was a central material for realizing semiconductor spin devices, such as the Das-Datta type spin transistor [1]. Recently, Si is the other pivotal material in semiconductor spintronics, since Si is a light element with lattice inversion symmetry and is ubiquitous and non-toxic. Especially, the lattice inversion symmetry and small atomic number of Si give rise to good spin coherence, enabling a novel spin device such as the Sugahara-Tanaka type spin metal-oxide-semiconductor field-effect transistor (MOSFET) [2]. Si spin MOSFET allows constructing for example a reconfigurable logic circuit, resulting in ultra-low power consumption logic systems. Hence, room-temperature (RT) operation of Si spin MOSFET has been a significant milestone in Si spintronics. In order to realize Si spin MOSFET operating at RT, the shortest cut is to realize electrical spin injection into Si at RT, and much effort has been paid. A variety of experimental methods for obtaining steadfast evidence of successful electrical spin injection in Si, and it is now widely recognized that a combination of a non-local 4-terminal method [3,4], local methods including the 3-terminal-magnetoresistance (3T-MR) [5] and 2-terminal [6] methods, and the Hanle effect measurement [3,7] is indispensable in order to avoid detection of spurious signals in the conventional local 3-terminal method [8-12]. The first reliable spin transport in Si at RT was reported in degenerate n-type Si [7], where the non-local 4-terminal method was used for propagation of pure spin current and the Hanle-type spin precession was nicely observed up to RT. However, too high doping concentration of the degenerate Si impeded an effective gate-voltage application, i.e., a spin MOSFET operation. In 2014, the first experimental demonstration of the RT operation of Si spin MOSFET was achieved by using non-degenerate n-type Si [13]. Since the Si was not degenerated, the gate voltage application allowed the gate-induced modulation of a spin signal and the source-drain current, $I_{sd}$, simultaneously. However, the central remaining issue to be solved for a practical application was low on/off ratio of the device in both a spin signal and the source-drain current, $I_{sd}$. In this paper, we report on a Si spin MOSFET operation at RT with the high on/off ratio. The electrical local 2-terminal method was used for investigating the spin MOSFET characteristics, where the spin transport was corroborated by the 3T-MR method. The on/off ratio of the $I_{sd}$ and spin signals is greater than $10^3$ in the same Si spin MOSFET. More significantly, the gate voltage dependence of the spin signals is consistent with that of the $I_{sd}$. This achievement can pave a way to a practical application of a Si spin MOSFET.

The Si spin MOSFET was fabricated on a silicon-on-insulator (SOI) substrate with a structure of Si(100 nm)/$SiO_2$(200 nm)/bulk Si (see Fig. 1(a)). The upper Si layer was phosphorous (P) doped by ion implantation. Before depositing the ferromagnetic electrodes, the Si in these regions was highly-doped to a concentration of about $5 \times 10^{19}$ $cm^{-3}$. The doping profile measured by secondary ion mass spectroscopy is shown in Fig. 1(b). The profile is the same as that of the spin MOSFET investigated in the literature [13], because we used the same batch of the SOI substrate. After the natural oxide layer on the Si channel was removed using a HF solution, Pd(3 nm)/Fe(13 nm)/MgO(0.8 nm) was grown on the etched surface by molecular beam epitaxy. Then, we etched out the Pd (3 nm)/Fe (3 nm) layers and Ta (3nm) was grown on the remaining Fe. In order to realize an efficient gate voltage application, the surface of the Si channel was etched to a depth of more than 25 nm by ion milling. The device resistivity was 4300 Ω μm, which is much larger than that in the previously investigated spin MOSFET (160 Ω μm [13]), indicating that the etching process effectively works to remove the highly doped region at the surface of the Si channel. The contacts had dimensions of 0.5×21 μm$^2$ and 2×21 μm$^2$, respectively. The Si channel surface and sidewalls at the FM contacts were buried by $SiO_2$. The nonmagnetic electrodes, with dimensions of 21×21 μm$^2$, were made from Al and were produced by ion milling. The gap between the FM electrodes was set to be 4.26 μm. The gate voltage was applied from the backside of the device. A probing station (Janis, ST-500), a source meter (KEITHLEY 2400 and 2401) and a digital multimeter (KEITHLEY 2010) were used for investigating a conventional FET and spin MOSFET characteristics. All measurement was performed at RT.

Fig. 2 shows a principle of detecting spin signals in the 3T-MR and 2-terminal methods used in this study. As reported in the literature [5], the 3T-MR method is another reliable method for spin transport (not only spin accumulation), and resistance hysteresis due to spin transport can be observed, which is the central difference from the conventional local 3-terminal method. The ferromagnetic electrode as a spin detector was set in the downstream of the spin flow. Since the spin flow in the Si is affected by the spin drift effect due to an electric field in the Si [14,15], efficient spin accumulation in the downstream side is realized. Since the spin direction in two ferromagnetic electrodes is controlled by an external magnetic field parallel to the Si channel, i.e., an alignment of the spin direction of the detector electrode and the accumulated spin beneath the detector electrode (=the propagating spin in the Si) are controlled to be parallel or antiparallel, yielding the resistance hysteresis. The more quantitative and detailed discussion will be implemented in elsewhere [16]. On the contrary, spin accumulation at both spin injector and detector electrodes can be measured in the local 2-terminal method. Notable is that anisotropic magnetoresistance and other spurious signals can be superimposed in detected signals in the local method, careful experiments are necessary. We used the both methods for corroborating spin transport in the spin MOSFET at RT.

Fig. 3(a) shows the magnetoresistance including minor loop observed in the 3T-MR method. The magnitude of the magnetoresistance was measured to be 0.33 Ω when the electric current was set to be 1 mA. The magnetic fields, where the hysteresis and the minor loop appeared, is consistent with each other, indicating that the magnetoresistance is ascribed to successful spin transport in the Si spin MOSFET at RT. Since the distance of the ferromagnetic electrodes was set to be 4.26 μm, the spin transport of 4.26 μm was realized. Such long spin transport was achieved by spin drift in addition to spin diffusion as discussed previously [13]. When we changed the measuring method to the local 2-terminal method with the gate voltage application, the similar magnetoresistance was observed under an application of the gate voltage of +50 V as shown in Fig. 3(b) (note that the electric current instead of the device resistance is shown). As described in the literature [2], on- and off-states are generated by the gate voltage as a conventional MOSFET (electric on/off states), and in addition, the other on- and off-states are generated by the magnetization configuration (spin on/off states). We define a spin signal of the Si spin MOSFET as a difference of the $I_{sd}$ under the parallel and antiparallel magnetization configurations, i.e., in the spin on/off states as shown in Fig. 3(c). Under the gate voltage of +50 V, the magnitude of the spin signal was measured to be 0.68 μA.

When the gate voltage was swept from −125 to +125 V under the source-drain voltage of 4 V, the $I_{sd}$ of the spin MOSFET was modulated (see Fig. 4). Since the Si is n-type, the positive gate voltage gives rise to the larger $I_{sd}$, and the negative gate voltage suppresses the $I_{sd}$. Hence, a conventional MOSFET operation is achieved in the device. The switching voltage of the device is comparatively large; in other words, the transconductance of the device is not good enough. In addition, the minimum of the $I_{sd}$ can be seen at the gate voltage of -35 V. These are not expected in commercialized Si MOSFETs, and we deduce that they are attributed to a non-optimized device structure, and especially, the back gate structure can be a reason for them. Although the FET characteristics does not reach that of the commercialized ones, the electric on/off states are sufficiently separated, which meets a purpose of our experiments. To note is that the spin signals are also effectively modulated by the gate voltage (see Fig. 4). The magnitude of the spin signals was changed from ca. 1 μA under positive gate voltages to ca. 0.5 nA under negative gate voltages. The decrease of the signal is attributed to the following reason: when the gate voltage is smaller than ca. 50 V, the $I_{sd}$ starts to be suppressed. Here, the suppression equally takes place under the parallel and antiparallel magnetization alignments. When the $I_{sd}$ is suppressed, the difference of the $I_{sd}$ between the spin on/off states becomes smaller, resulting in the suppression of the spin signals. Since 0.5 nA is almost the detection limit in this measurement, it can be said that the spin signals were modulated over three orders of magnitude. The other significance is that the $I_{sd}$ and the magnitude of the spin signals were nicely consistent as a function of the gate voltage and that the on/off ratio of the $I_{sd}$ and the spin signals was greater than $10^3$ for both. Since the on/off ratio was much less than $10^1$ in the previous study [13], the optimization of the device fabrication process effectively improved the spin MOSFET performance. This result directly indicates the successful RT operation of the Si spin MOSFET.

In summary, we demonstrated the RT operation of the Si spin MOSFET with large on/off ratio of the $I_{sd}$ and the spin signals. Spin transport in non-degenerate Si at RT and an effective application of the gate voltage enabled the operation. This success can pave a way to practical applications of Si spin MOSFETs.

The authors thank Dr. T. Sasaki in TDK Corporation for his fruitful discussion. A part of this study performed by M.S. and Y.A. was supported by the Japan Society for the Promotion of Science (Kakenhi, Scientific Research (A)).

## Figures and figure captions

**Fig. 1(a)** Device structure of the Si spin MOSFET. **(b)** The doping profile of the Si spin channel measured by the secondary ion mass spectroscopy. The dashed line shows the position of the etching depth of 25 nm, where the phosphorous concentration was estimated to be $6\times10^{18}$ $cm^{-3}$.

**Fig. 2** Schematic images of the measurement methods in this study. The left and right images show the measurement circuit and a corresponding position dependence of an electrochemical potential in the local 3T-MR method and local 2-terminal method (2T). "Spin accumulation" in the lower panels show the spin accumulation voltages detected in each measurement method.

**Fig. 3(a)** Magnetoresistance (the upper panel) and minor loop (the lower panel) signals of the Si spin MOSFET in the local 3T-MR method. The external magnetic field was applied along the y-direction. **(b)** Magnetoresistance of the Si spin MOSFET in the local 2-terminal method with the application of the gate voltage ($V_g$) of +50 V. The source-drain current ($V_{sd}$) was set to be 4 V. The external magnetic field was applied along the y-direction. The magnitude of the spin signals was measured to be 0.68 μA. The detail is described in the main text. **(c)** A schematic image of the spin on/off state in the Si spin MOSFET.

**Fig. 4** Device characteristics of the Si spin MOSFET. The left and right vertical axes show the source-drain current, $I_{sd}$, and the spin signal (the difference of the $I_{sd}$ under parallel (P) and antiparallel (AP) configurations). The black solid line and the red closed circles show the gate

voltage dependence of the the $I_{sd}$ and the spin signals, respectively. The source-drain current ($V_{sd}$) was set to be 4 V

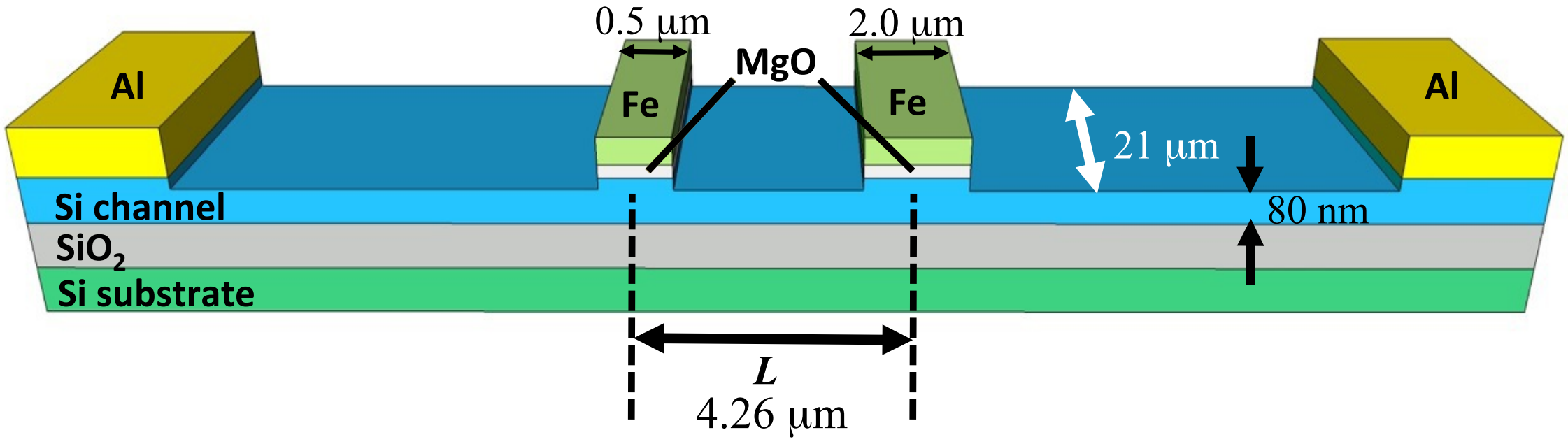

**Fig. 1(a) T. Tahara et al.**

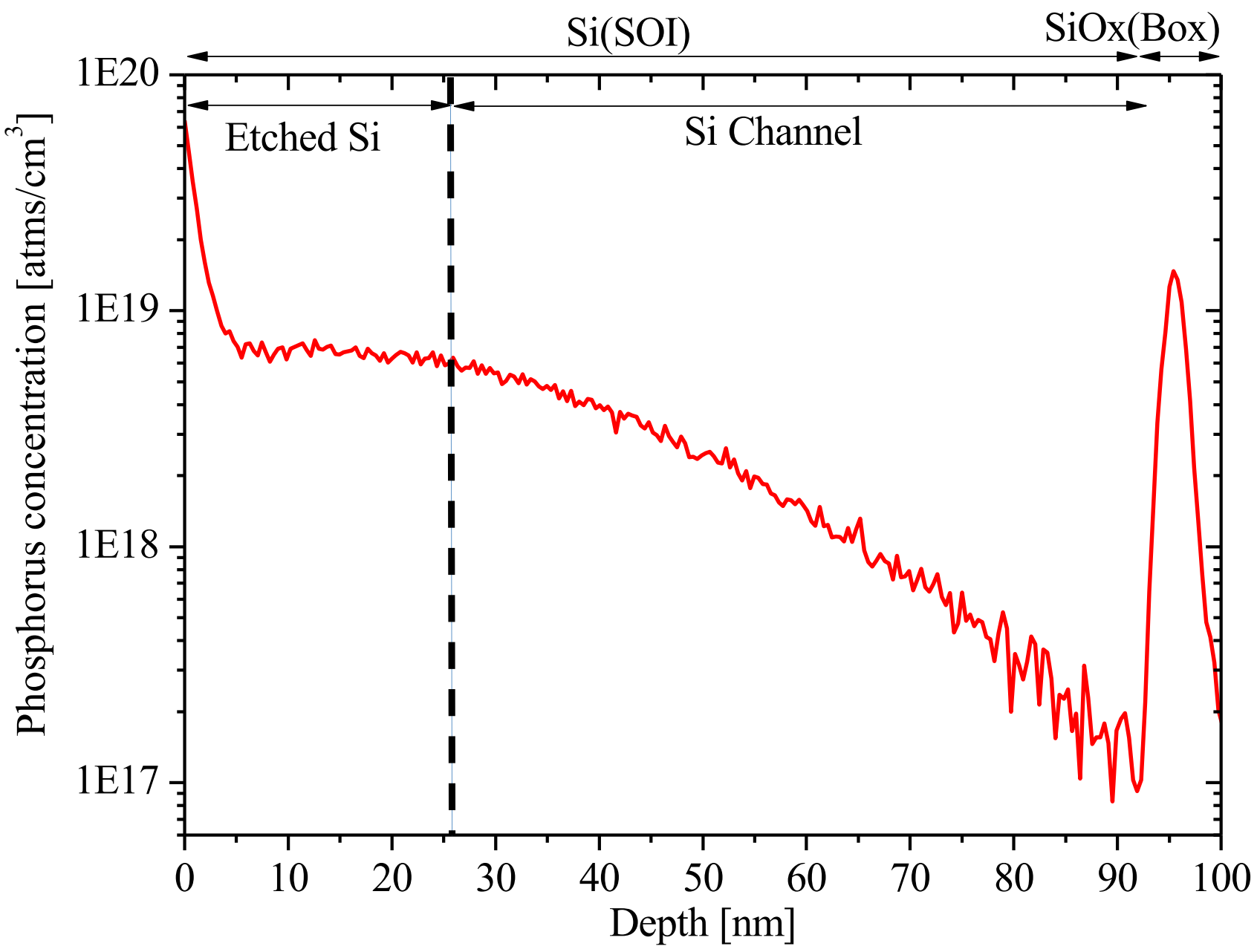

**Fig. 1(b) T. Tahara et al.**

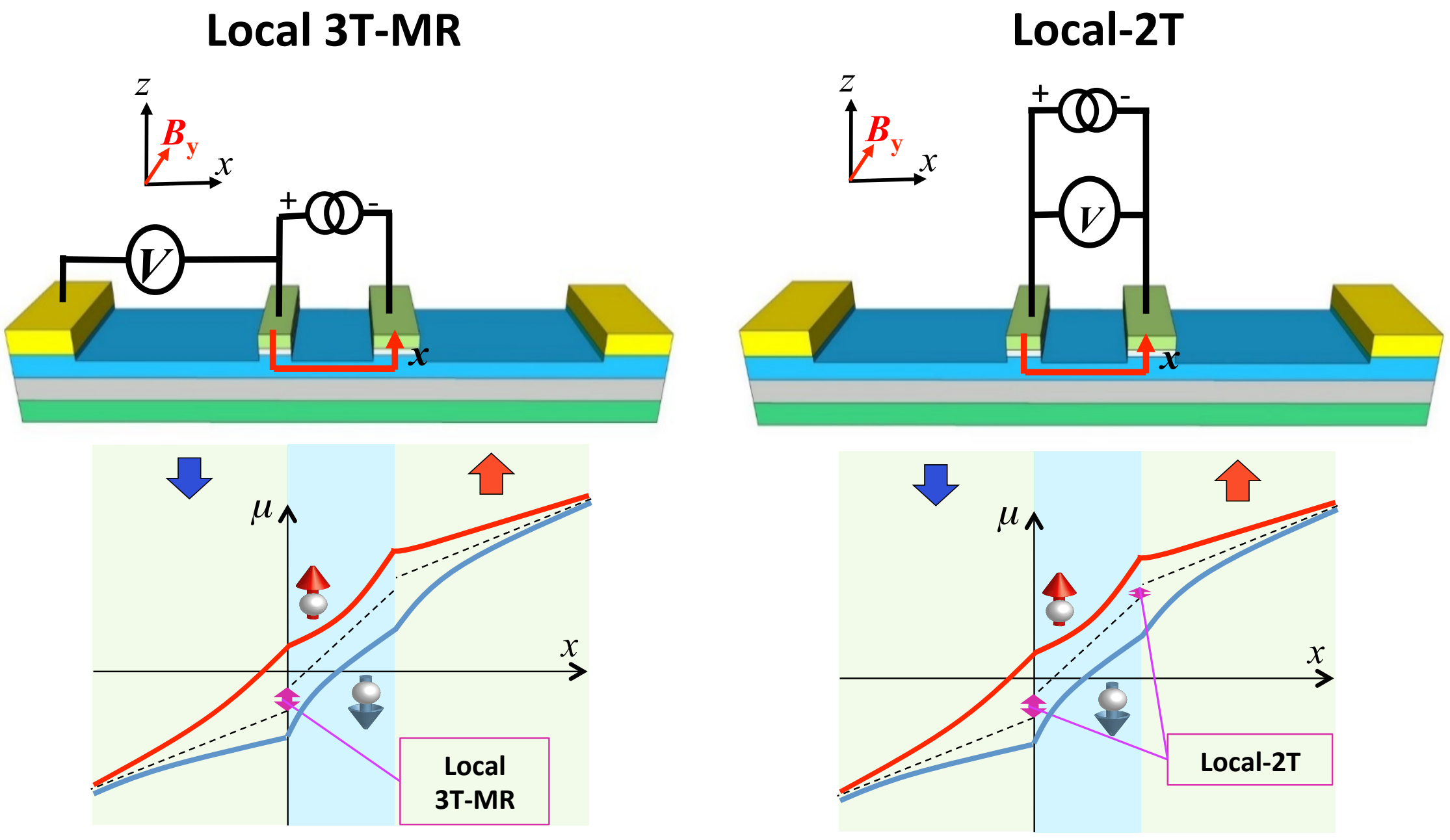

**Fig. 2 T. Tahara et al.**

(a)

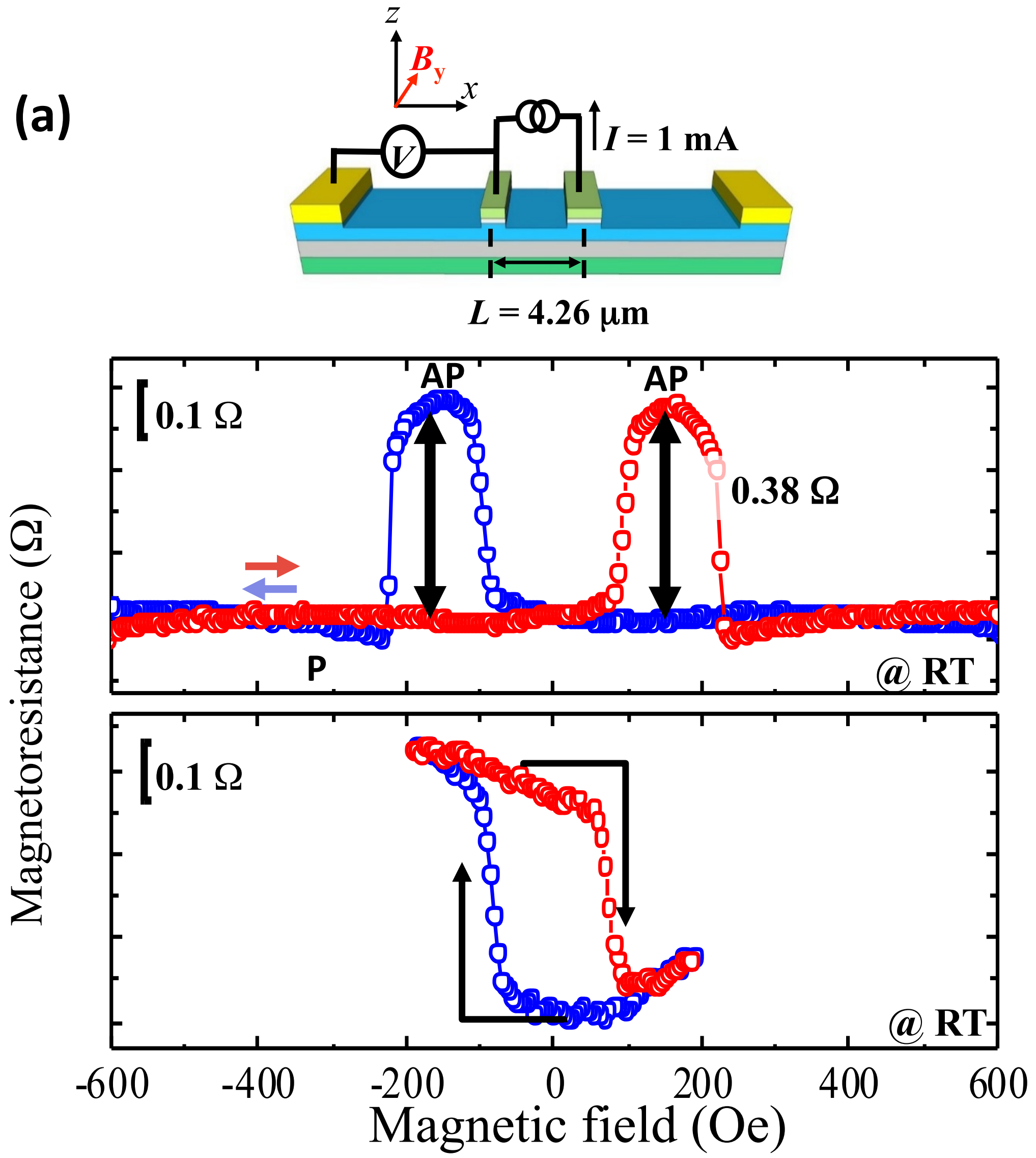

**Fig. 3(a) T. Tahara et al.**

(b)

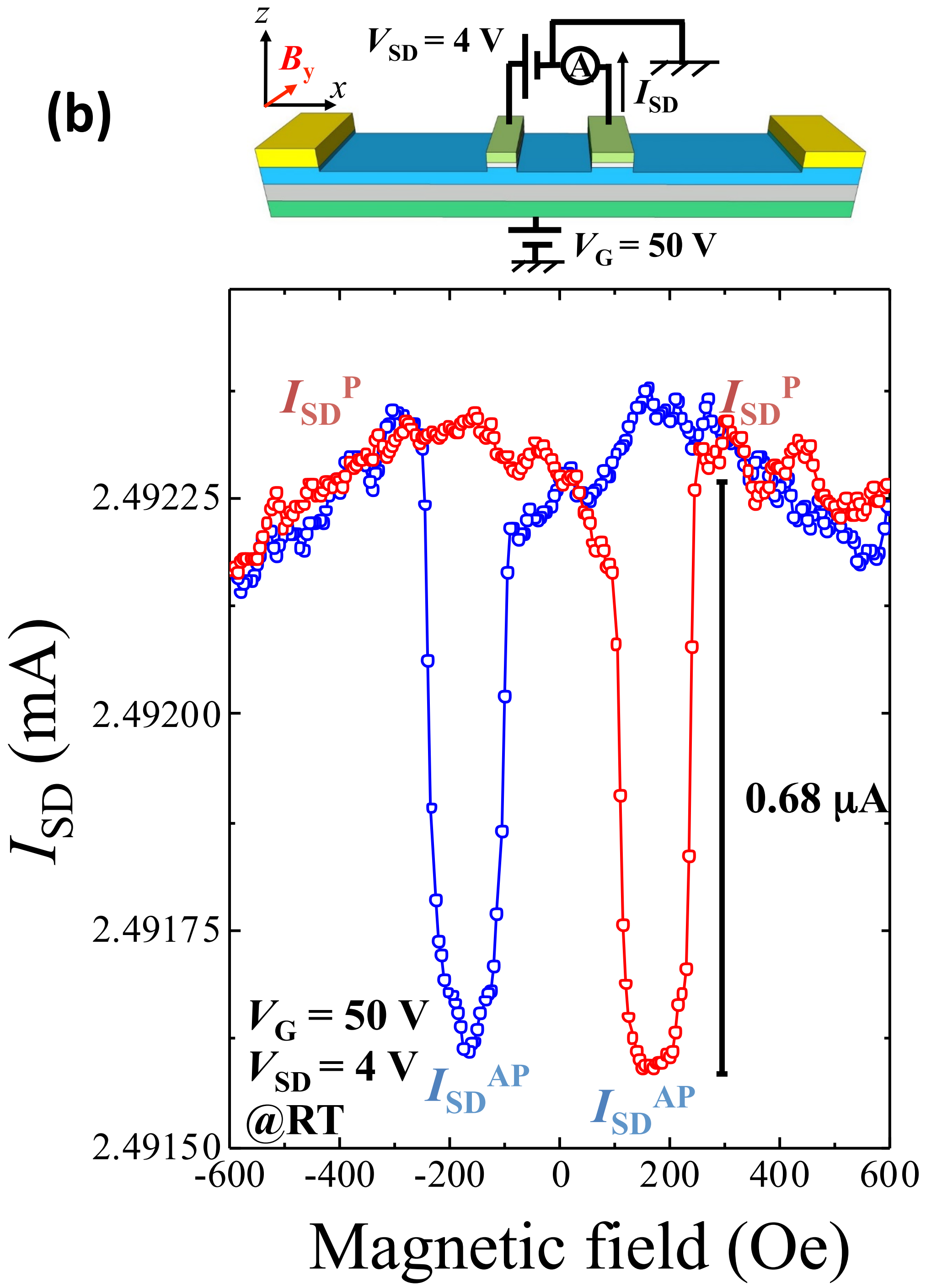

Fig. 3(b) T. Tahara et al.

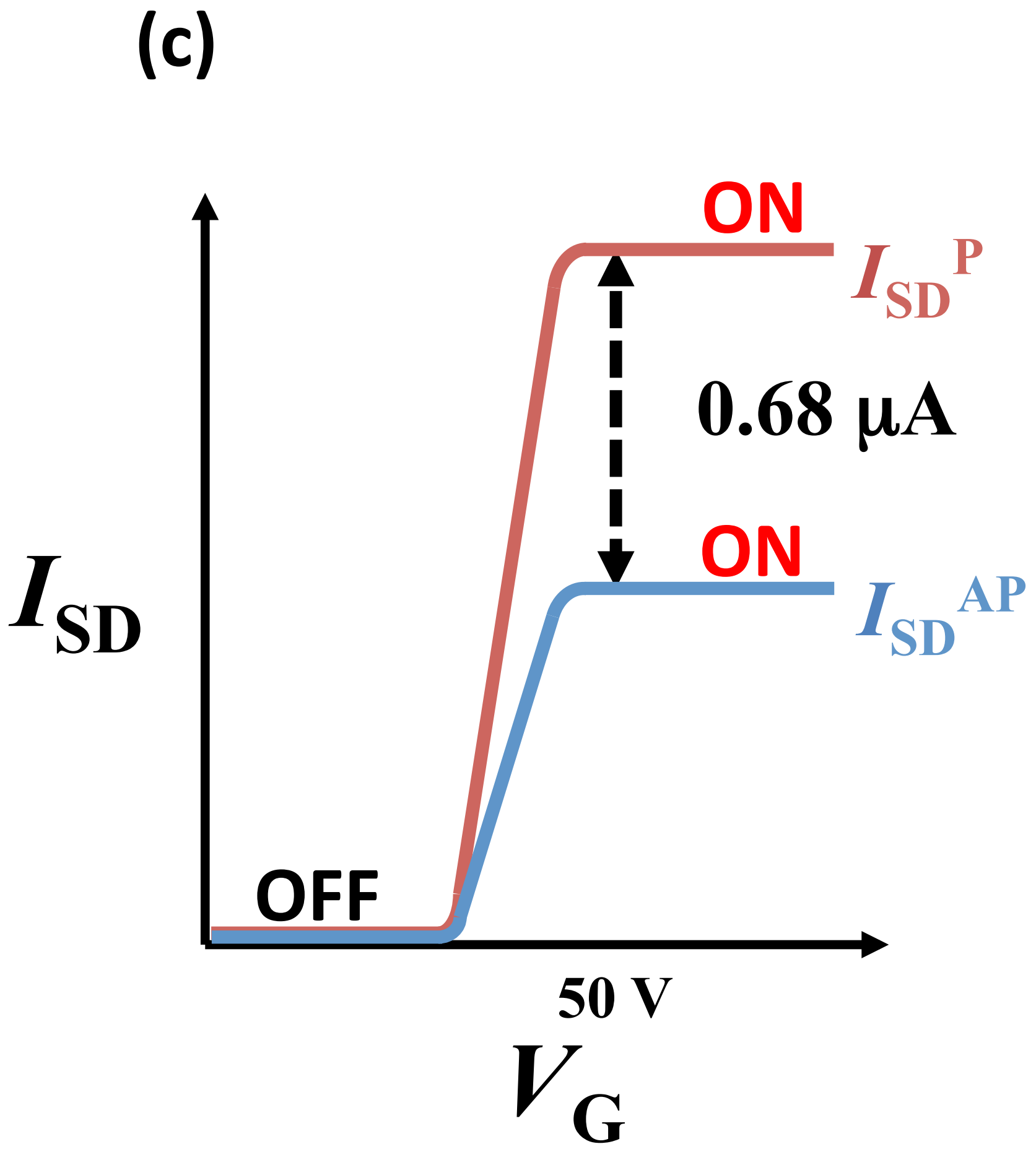

Fig. 3(c) T. Tahara et al.

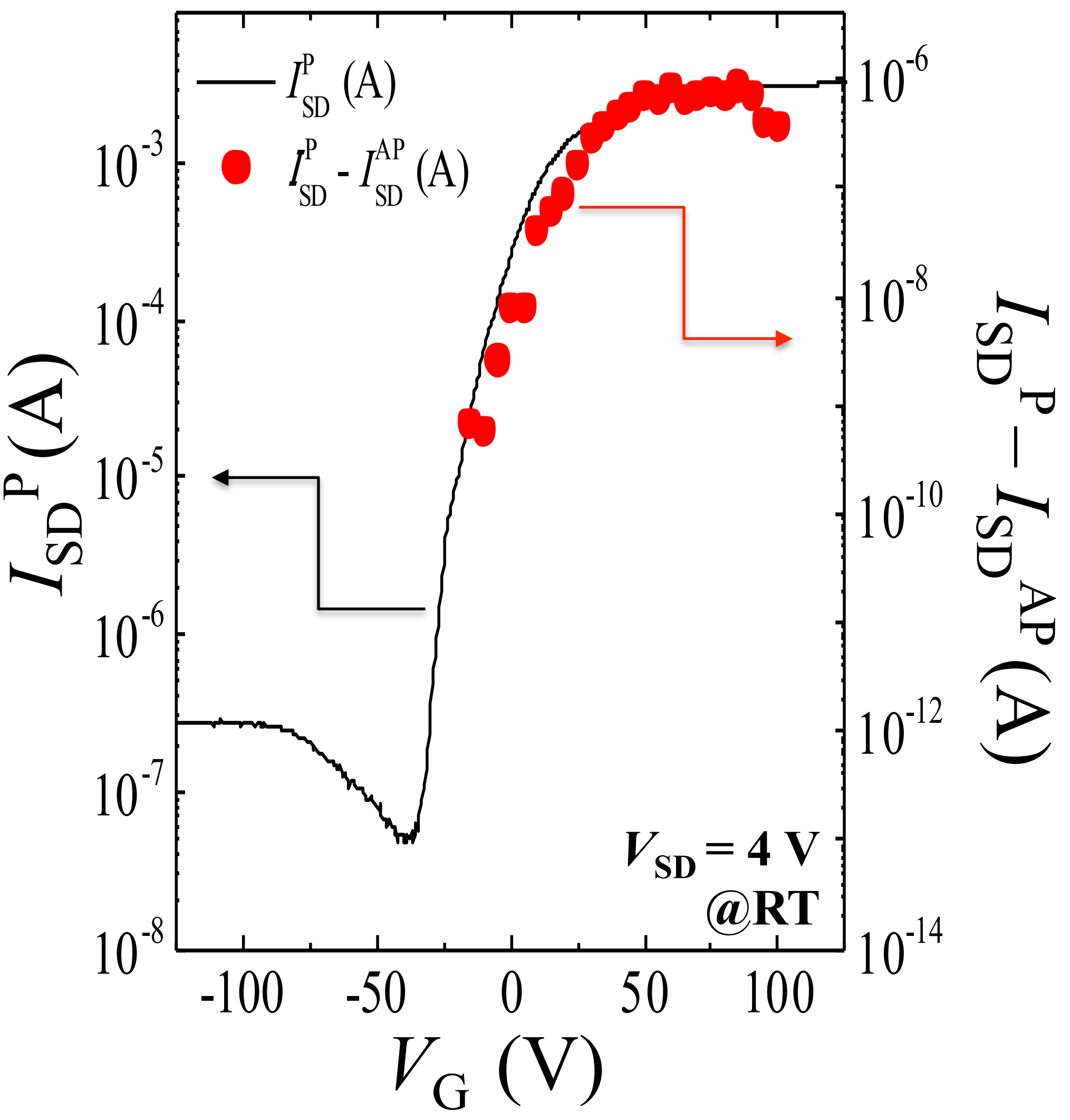

**Fig. 4 T. Tahara et al.**